\documentclass[letterpaper, 10 pt, conference]{ieeeconf}  

\IEEEoverridecommandlockouts                              

\usepackage{cite}
\usepackage{amsmath,amssymb,amsfonts}
\usepackage{algorithmic}
\usepackage{graphicx}
\usepackage{algorithm,algorithmic}
\usepackage{hyperref}
\hypersetup{hidelinks=true}
\usepackage{textcomp}
\usepackage{graphicx}
\usepackage{epsfig}  
\usepackage{amsmath}
\usepackage{indentfirst}  
\usepackage{color}
\usepackage[dvipsnames]{xcolor}
\usepackage{subcaption}
\usepackage{tikz}
\usepackage{soul}
\usepackage{xcolor}
\usepackage{mathtools}
\usepackage{optidef}
\usepackage{breqn}
\usepackage[algo2e,norelsize]{algorithm2e} 

\SetKwInput{kwAdd}{Add}
\SetKwInput{kwSet}{Set}
\SetKwInput{kwReset}{Reset}
\SetKwInput{kwUpdate}{Update}
\SetKwInput{kwClear}{Clear}
\SetKwInput{kwSave}{Save}
\SetKwInput{kwBuild}{Build}
\SetKwInput{kwIni}{Initialization}
\usetikzlibrary{matrix,chains,positioning,decorations.pathreplacing,arrows}

\newtheorem{remark}{Remark}

\newtheorem{assumption}{Assumption}

\newcommand{\bG}{{\mathbf G}}

\newcommand{\bL}{{\mathbf L}}

\title{\LARGE \bf Distributed Koopman Learning with Incomplete Measurements \thanks{}
}

\author{Wenjian Hao, Lili Wang, Ayush Rai, Shaoshuai Mou 
\thanks{This material is based upon work supported by the Defense Advanced Research Projects Agency (DARPA) under of the Learning Introspective Control (LINC) project (grant no. N65236-23-C-8012) Any opinions, ﬁndings and conclusions or recommendations expressed in this material are those of the author(s) and do not necessarily reﬂect the views of the DARPA or the U.S. Government.}
\thanks{W. Hao, L. Wang, A. Rai, and S. Mou are with the School of Aeronautics and Astronautics at Purdue University. Email: {\tt\small \{hao93, wang6127, rai29, mous\}@purdue.edu}.}}

\begin{document}

\maketitle
\thispagestyle{empty}
\pagestyle{empty}

\begin{abstract}
Koopman operator theory has emerged as a powerful tool for system identification, particularly for approximating nonlinear time-invariant systems (NTIS). This paper considers a network of agents with limited observation capabilities that collaboratively estimate the dynamics of an NTIS. A distributed deep Koopman learning algorithm is developed by integrating Koopman operator theory, deep neural networks, and consensus-based coordination. In the proposed framework, each agent approximates the system dynamics using its partial measurements and lifted states exchanged with its neighbors. This cooperative scheme enables accurate reconstruction of the global dynamics despite the absence of full-state information at individual agents. Simulation results on the Lunar Lander environment from OpenAI Gym demonstrate that the proposed method achieves performance comparable to the centralized deep Koopman learning with full-state access.
\end{abstract}

\section{Introduction}
\label{sec:introduction}
Many applications of model-based optimal control design, such as model predictive control and linear-quadratic regulator, rely on the availability of accurate dynamics models for state propagation. With the growth in complexity of autonomous systems in terms of their dynamics \cite{ mamakoukas2021derivative, mezic2015applications, proctor2018generalizing, mauroy2016linear}, there has been a surge of interest in learning these dynamics through data-driven methods. 

One popular approach involves leveraging Koopman operator theory, which can approximate a nonlinear system in a linear form based on the available data in the form of state-control pairs \cite{mezic2015applications, proctor2018generalizing, mauroy2016linear}. The fundamental idea is to lift the state space to a higher-dimensional space with approximately linear dynamics evolution. Early methods, such as dynamic mode decomposition (DMD) \cite{tu2013dynamic, kutz2016dynamic} and extending dynamic mode decomposition (EDMD) \cite{korda2018linear} implement this idea but require manual specification of observable functions, often resulting in excessively high-dimensional lifted spaces that hinder real-time implementation. To address this issue, several extensions have been developed that automate the choice of observables. For example, \cite{lusch2017data} introduces deep learning techniques to discover Koopman eigenfunctions, while subsequent works \cite{yeung2019learning, dk2, dk, bevanda2021koopmanizingflows} utilize deep neural networks (DNNs) to parameterize observables and optimize them via loss minimization, a class of methods referred to as deep Koopman operator (DKO) approaches. More recently, \cite{hao2024deep} extends DKO methods to handle nonlinear time-varying systems.

As multi-agent systems (MAS) become increasingly prevalent in large-scale applications, there arises a need for distributed frameworks capable of identifying complex dynamical systems. In particular, we focus on scenarios where each agent has access only to partial observations of a nonlinear time-invariant system (NTIS) and must cooperatively estimate the system dynamics in a distributed manner. This challenge is fundamentally different from distributed state estimation problems \cite{olfati2005consensus,park2016design, wang2017distributed}, where the dynamics are assumed to be known and the objective is to reconstruct the full system state. Instead, our goal is to capture the full dynamical evolution of the system itself.

Prior work on distributed dynamics learning has often relied on parametric modeling, where the system is characterized by unknown parameters and the task is reformulated as distributed estimation. For instance, diffsion-based adaptive methods were proposed in \cite{lopes2008diffusion, cattivelli2009diffusion} to achieve least-mean-square solutions, while \cite{fu2022distributed} addressed distributed identification of linear stochastic systems with binary sensors and partially unknown coefficients. However, these parameter-based formulations require prior knowledge of model structure, which is not always available. Our interest lies in estimating the dynamics directly through linearized approximations, thereby enabling downstream tasks such as controller design and optimal control.

Motivated by these challenges, this paper proposes a distributed deep Koopman learning framework for approximating NTIS. In contrast to centralized Koopman operator approaches \cite{nandanoori2021data, mukherjee2022learning}, which construct block-wise global models, the proposed method achieves a fully distributed approximation by leveraging consensus among agents. Each agent operates with only partial measurements and communicates solely with its neighbors. During training, agents exchange estimated full-state representations and lifted Koopman observables, while during prediction, only neighboring lifted observables are required. This design significantly reduces both communication and computational demands, and improves learning resiliency.

This paper is structured as follows: Section \ref{pformulation} formulates the problem. Section \ref{pmethod} presents the proposed algorithm. Section \ref{results} provides numerical simulations to demonstrate the effectiveness of the approach. Finally, Section \ref{conclusion} concludes the paper with a summary of key findings.

\textbf{\emph{Notations.}} The Euclidean norm of a vector is denoted by $\parallel \cdot \parallel$. For a matrix $A\in\mathbb{R}^{n\times m}$, $\parallel A \parallel_F$ denotes its Frobenius norm, $A'$ denotes its transpose, and $A^\dagger$ denotes its Moore-Penrose pseudoinverse. 

\section{The Problem}\label{pformulation}
Consider the following discrete-time nonlinear time-invariant system (NTIS):
\begin{equation}\label{eq_NTIS}
    \boldsymbol{x}(t+1) = \boldsymbol{f}(\boldsymbol{x}(t), \boldsymbol{u}(t)),
\end{equation}
where $t=0,1,2,\cdots$ denotes the time index, $\boldsymbol{x}(t)\in \mathbb{R}^n$ and $\boldsymbol{u}(t) \in \mathbb{R}^m$ denote the system state and control input at time $t$, respectively, and $\boldsymbol{f}:\mathbb{R}^n\times\mathbb{R}^m\rightarrow\mathbb{R}^n$ is the time-invariant and unknown dynamics mapping. A state–input trajectory generated by $\boldsymbol{f}$ from time $0$ to $T$ is given by:\begin{equation}\label{eq_traj_full}
    \boldsymbol{\xi}=\{(\boldsymbol{x}_t,\boldsymbol{u}_t):  t=0,1,2,\cdots,T\}.  
\end{equation} Here, $(\boldsymbol{x}_t,\boldsymbol{u}_t)$ denotes an observed constant state–input pair, distinguishing it from the state–input variables $(\boldsymbol{x}(t),\boldsymbol{u}(t))$ in the rest of this paper.

One way to approximate the unknown dynamics $\boldsymbol{f}$ in \eqref{eq_NTIS} using $\boldsymbol{\xi}$ is the DKO method, which aims to find constant matrices $\hat A^*\in\mathbb{R}^{r\times r}$, $\hat B^*\in\mathbb{R}^{r\times m}$, $\hat H^*\in\mathbb{R}^{n\times r}$, and a parameter vector $\boldsymbol{\hat \theta}^*\in\mathbb{R}^p$ such that for any $0\leq t\leq T-1$, the following holds:
\begin{equation}\label{eq_DKO}
    \boldsymbol{x}_{t+1} = \hat H^*\Big(\hat A^* \boldsymbol{g}(\boldsymbol{x}_t,\boldsymbol{\hat \theta}^*) + \hat B^* \boldsymbol{u}_t\Big),
\end{equation} where $\boldsymbol{g}(\cdot,\boldsymbol{\hat \theta}^*):\mathbb{R}^n\rightarrow\mathbb{R}^r$ is a known function, and the lifting dimension satisfies $r\geq n$.

Consider a group of $N\geq 1$ agents, with the node set denoted as $\mathcal{V} = \{1,2,\dots, N\}$. Each agent $i \in \mathcal{V}$ can receive information from its neighbors. By saying neighbors of agent $i$, we mean any other agents within agent $i$'s reception range, and we denote $\mathcal{N}_i$ as the set of agent $i$'s neighbors. Neighbor relations between distinct pairs of agents are characterized by a self-arced directed graph $\mathbb{G} = \{\mathcal{V},\mathcal{E}\}$ such that there is a directed edge $(i,j)\in\mathcal{E}$ if and only if agents $i$ and $j$ are neighbors. We assume $\mathbb{G}$ is strongly connected and note that each agent $i$ is always considered a neighbor of itself. 

\begin{figure}[h]
    \centering
\includegraphics[width=0.8\linewidth]{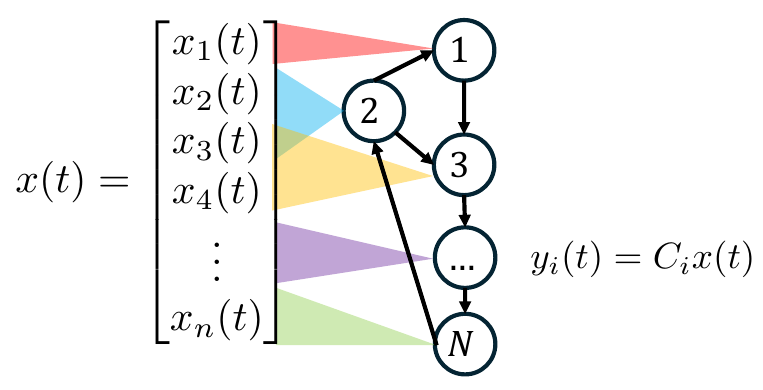}
    \caption{Illustration of the partial observations.}
    \label{fig:illustration}
\end{figure}

As shown in Fig. \ref{fig:illustration}, at any time $t$, each agent $i$ in the multi-agent systems (MAS) shares the same control inputs $\boldsymbol{u}(t)$ and simultaneously observes a partial system state $\boldsymbol{y}_i(t)\in\mathbb{R}^{n_i}$, given by: \begin{equation}
    \boldsymbol{y}_i(t) = C_i \boldsymbol{x}(t), \nonumber
\end{equation} where $C_i\in\mathbb{R}^{n_i\times n}$ is the observation matrix of agent $i$. Correspondingly, all agents in the MAS synchronously construct the partial measurement–input trajectories of $\boldsymbol{\xi}$ in \eqref{eq_traj_full}:  \begin{equation}\label{eq_batch}
    \boldsymbol{\xi}_i=\{(\boldsymbol{y}_{i,t},\boldsymbol{u}_t):  t=0,1,2,\cdots,T\},
\end{equation} where $\boldsymbol{y}_{i,t}=C_i\boldsymbol{x}_t$ denotes the observed constant state vector at time $t$, in contrast to the partial state variable $\boldsymbol{y}_i(t)$. Note that $\boldsymbol{\xi}_i$ shares the same time index as the complete state trajectory $\boldsymbol{\xi}$ in \eqref{eq_traj_full}. To guarantee that $\boldsymbol{x}(t)$ can be uniquely recovered from $\boldsymbol{y}_i(t)$, we impose the following assumption.
\begin{assumption}
Let $C = [C_1', C_2', \cdots, C_N']'\in\mathbb{R}^{\bar n \times n}$ with $\bar n = \sum_{i=1}^N n_i$. We assume that the matrix $C \in\mathbb{R}^{\bar n\times n}$ has full column rank $n$, ensuring that the system is jointly observable. 
    \label{assumption1}
\end{assumption}

In the considered setting, at any time $t$, each agent $i$ in the MAS observes only the partial state $\boldsymbol{y}_i(t)$, rather than the full state $\boldsymbol{x}(t)$. Such local information is typically insufficient for independently identifying the dynamics in \eqref{eq_DKO}. 

To address this limitation, the \textbf{problem of interest} for each agent $i$ is to design a distributed algorithm that identifies constant matrices $A_i^*\in\mathbb{R}^{r\times r}$, $B_i^*\in\mathbb{R}^{r\times m}$, $H_i^*\in\mathbb{R}^{n\times r}$, and a parameter vector $\boldsymbol{\theta}_i^*\in\mathbb{R}^p$ such that, for any $0\leq t\leq T-1$, 
 \begin{equation}\label{eq_DDKO}
    \boldsymbol{x}_{t+1} = H_i^*\Big(A_i^* \sum_{j\in\mathcal{N}_i}\boldsymbol{g}_j(\boldsymbol{y}_{j,t},\boldsymbol{\theta}_j^*) + B_i^* \boldsymbol{u}_t\Big),
\end{equation}
where $\boldsymbol{g}_i(\cdot, \boldsymbol{\theta}_i):\mathbb{R}^{n_i}\rightarrow\mathbb{R}^r$ is a known function parameterized by $\boldsymbol{\theta}_i\in\mathbb{R}^{p_i}$ with $r\geq n$. The proposed updates for achieving \eqref{eq_DDKO} require each agent $i$ to exchange the constant vectors $\boldsymbol{y}_{i,t}$ and $\boldsymbol{g}_i(\boldsymbol{y}_{i,t}, \boldsymbol{\theta}_i)$ with its neighbors.

The problem considered in this paper is thus a distributed data-fitting problem, where each agent $i$ uses its local dataset $\boldsymbol{\xi}_i$ and neighbor information to learn a mapping from $(\boldsymbol{y}_{i,t}, \boldsymbol{u}_t)$ to $\boldsymbol{x}_{t+1}$. As shown in Section~\ref{results}, the dynamics in \eqref{eq_DDKO} identified by the proposed algorithm achieve accuracy comparable to the centralized DKO in \eqref{eq_DKO}.
\begin{remark}
    An alternative to \eqref{eq_DKO} is to first estimate the full state and then reformulate the multi-agent problem as a centralized one. However, estimating the complete system state may require an infinite number of time steps, thereby degrading the performance of the learned dynamics model in deployment. In contrast, the dynamics in \eqref{eq_DDKO} rely only on $\boldsymbol{y}_{i,t}$, $\boldsymbol{u}_t$, and the lifted states $\sum_{j\in\mathcal{N}_i}\boldsymbol{g}_j(\boldsymbol{y}_{j,t},\boldsymbol{\theta}_j)$, $j\neq i$ from its neighbors for prediction.
\end{remark}

\section{Main Results}\label{pmethod}
In this section, we first outline the main challenges and key concepts of the proposed problem, and then present a distributed algorithm to achieve \eqref{eq_DDKO}.

\subsection{Challenges and Key Ideas}
Consider an arbitrary agent $i$ in the MAS. We define $$\boldsymbol{z}_{i, t} = \sum_{j\in\mathcal{N}_i}\boldsymbol{g}_j(\boldsymbol{y}_{j,t},\boldsymbol{\theta}_j),$$ where $\boldsymbol{g}_j(\boldsymbol{y}_{j,t},\boldsymbol{\theta}_j), j\neq i$ are the constant vectors received from its neighbors. To obtain \eqref{eq_DDKO}, we consider the following family of linear systems based on the Koopman operator theory:
\begin{align}
    \boldsymbol{z}_{i, t+1} &= A_i\boldsymbol{z}_{i, t} + B_i \boldsymbol{u}_t, \label{eqq1}  \\ \boldsymbol{x}_{t+1} &= H_i \boldsymbol{z}_{i, t+1}, \label{eqq2}
\end{align} 
where \eqref{eqq1} describes the system dynamics in the lifted space, and \eqref{eqq2} provides a linear mapping from the lifted state  $\boldsymbol{z}_{i, t+1}$ to the complete system state $\boldsymbol{x}_{t+1}$. To this end, we define the following optimization problem for each agent $i$ over $\boldsymbol{\xi}_i$:
\begin{equation}\label{eq_loss_single}
    \begin{aligned}
        \min_{A_i, B_i, H_i, \boldsymbol{\theta}_i} \frac{1}{2T}\sum_{t=0}^{T-1}\Big(\parallel \boldsymbol{z}_{i, t+1} - A_i\boldsymbol{z}_{i, t} - B_i \boldsymbol{u}_t \parallel^2 \\ + \parallel \boldsymbol{x}_{t+1} - H_i\boldsymbol{z}_{i, t+1} \parallel^2\Big),
    \end{aligned}
\end{equation} where the first and second parts of \eqref{eq_loss_single} are designed to approximate \eqref{eqq1} and \eqref{eqq2}, respectively.

A key challenge to solve \eqref{eq_loss_single} is that the complete state $\boldsymbol{x}_{t+1}$ is unavailable to any agent $i$. To overcome this, we propose an iterative algorithm in which each agent simultaneously estimates the complete state and minimizes \eqref{eq_loss_single}.

\subsection{Algorithm}
We now present a distributed algorithm that solves \eqref{eq_loss_single}. To begin, for each agent $i$, we construct the following data matrices from its trajectory $\boldsymbol{\xi}_i$ in \eqref{eq_batch}:
\begin{equation}\label{eq_data_mat}
    \begin{aligned}
    \mathbf{\bar X} &=[\boldsymbol{x}_1, \boldsymbol{x}_2,\cdots,\boldsymbol{x}_T]\in \mathbb{R}^{n \times T},  \\
    \mathbf{\bar Y}_i &= [\boldsymbol{y}_{i,1}, \boldsymbol{y}_{i,2},\cdots,\boldsymbol{y}_{i,T}]\in \mathbb{R}^{n_i \times T},\\
    \mathbf{U} &=[\boldsymbol{u}_0, \boldsymbol{u}_1,\cdots,\boldsymbol{u}_{T-1}]\in \mathbb{R}^{m \times T},\\
    \bG_i &= [\boldsymbol{g}_i(\boldsymbol{y}_{i,0},\boldsymbol{\theta}_i),\boldsymbol{g}_i(\boldsymbol{y}_{i,1},\boldsymbol{\theta}_i),\cdots, 
    \boldsymbol{g}_i(\boldsymbol{y}_{i,T-1}, \boldsymbol{\theta}_i)] \in \mathbb{R}^{r \times T},\\
    \mathbf{\bar G}_i &= [\boldsymbol{g}_i(\boldsymbol{y}_{i,1}, \boldsymbol{\theta}_i),\boldsymbol{g}_i(\boldsymbol{y}_{i,2},\boldsymbol{\theta}_i),\cdots, \boldsymbol{g}_i(\boldsymbol{y}_{i,T}, \boldsymbol{\theta}_i)] \in \mathbb{R}^{r \times T},\\ \mathbf{Z}_i &=\sum_{j\in\mathcal{N}_i}\mathbf{G}_j, \quad \mathbf{\bar Z}_i =\sum_{j\in\mathcal{N}_i}\mathbf{\bar G}_j.
    \end{aligned}
\end{equation}

\textbf{Full states estimation.} To estimate the complete system state matrix $\mathbf{\bar X}$, we formulate the following linear equation based on the definition of $\boldsymbol{y}_i(t) = C_i \boldsymbol{x}(t)$:
\begin{equation}\label{eq_dis_eq}
\begin{bmatrix}
    C_1 \\ C_2\\ \vdots \\C_N
\end{bmatrix} \mathbf{\bar X} = \begin{bmatrix}
    \mathbf{\bar Y}_1 \\  \mathbf{\bar Y}_2 \\ \vdots \\  \mathbf{\bar Y}_N
\end{bmatrix}.
\end{equation}
Recall $C = [C_1', C_2', \cdots, C_N']'$, and Assumption \ref{assumption1} ensures the existence of a unique $\mathbf{\bar X}$ that satisfies \eqref{eq_dis_eq}. Inspired by the work in \cite{mou2015distributed, wang2019distributed}, one distributed approach for each agent $i$ to achieve the $\mathbf{\bar X}$ in \eqref{eq_dis_eq} is to iteratively update its estimation matrix $\mathbf{\hat X}_i(k)$ with $C_i \mathbf{\hat X}_i(0)=\mathbf{\bar Y}_i$ as follows:
\begin{equation}\label{eq_Xmat_update}
        \mathbf{\hat X}_i(k+1) = \mathbf{\hat X}_i(k) + \frac{1}{d_i}(I_n- P_i)\sum_{j\in\mathcal{N}_i}(\mathbf{\hat X}_j(k) - \mathbf{\hat X}_i(k)),
\end{equation}
where $k=0,1,2,\cdots$ denotes the iteration index, and $P_i = C_i'(C_i C_i')^{-1}C_i$ denotes an orthogonal projection matrix, and $d_i$ denotes the number of neighbors of agent $i$.

\textbf{Dynamics learning.} Using notations in \eqref{eq_data_mat} and replacing the true system state matrix $\mathbf{\bar X}$ in \eqref{eq_loss_single} with the estimated complete state matrix $\mathbf{\hat X}_i$ from the update rule in \eqref{eq_Xmat_update}, we define the following loss function based on \eqref{eq_loss_single}:
\begin{equation}\label{eq_loss_compact}
\begin{aligned}
     &\bL(A_i, B_i, H_i, \boldsymbol{\theta}_i) \\=& \frac{1}{2T}\Big(\underbrace{\parallel \mathbf{\bar Z}_i- [A_i, B_i] \begin{bmatrix}
         \mathbf{Z}_i \\ \mathbf{U}
     \end{bmatrix}\parallel_F^2}_{\delta(A_i, B_i, \boldsymbol{\theta}_i)} +\underbrace{\parallel \mathbf{\hat X}_i - H_i\mathbf{\bar Z}_i \parallel_F^2}_{\bar\delta(C_i, \boldsymbol{\theta}_i)}\Big).
\end{aligned}
\end{equation} Let $A_i(k)$, $B_i(k)$, $H_i(k)$, $\boldsymbol{\theta}_i(k)$ denote the estimates of $A_i^*$, $B_i^*$, $H_i^*$, $\boldsymbol{\theta}_i^*$ in \eqref{eq_DDKO} at iteration $k$, respectively. Define $\bG_i(k)$ as $\bG_i$ in \eqref{eq_data_mat} with $\boldsymbol{\theta}_i(k)$, and $\mathbf{\bar Z}_i(k)$ similarly computed using $\boldsymbol{\theta}_i(k)$. 
If the matrices $\mathbf{\bar Z}_i(k) \in \mathbb{R}^{r\times T}$ and $\begin{bmatrix} \mathbf{\bar Z}_i(k)  \\ \mathbf{U} \end{bmatrix}\in \mathbb{R}^{(r+m)\times T}$ are full row rank (i.e., right-invertible), then after updating $\mathbf{\hat X}_i$ according to \eqref{eq_Xmat_update}, we adopt the following iterative update rule to minimize \eqref{eq_loss_compact}: 
\begin{equation}\label{lmn}
\small{[A_i(k), B_i(k)] = \arg\min_{[A_i, B_i]}\delta(A_i, B_i, \boldsymbol{\theta}_i(k)) = \mathbf{\bar Z}_i(k) \begin{bmatrix}
         \mathbf{Z}_i(k) \\ \mathbf{U}
     \end{bmatrix}^\dagger},
\end{equation}
\begin{equation}\label{lmn1}
    H_i(k) = \arg\min_{H_i}\bar\delta(C_i, \boldsymbol{\theta}_i(k)) = \mathbf{\hat X}_i(k+1)\mathbf{\bar Z}_i(k)^\dagger,
\end{equation}
\begin{equation}\label{lmn2}
    \boldsymbol{\theta}_i(k+1) = \boldsymbol{\theta}_i(k) - \alpha_i(k)\nabla_{\boldsymbol{\theta}_i}\bL(A_i(k), B_i(k), H_i(k), \boldsymbol{\theta}_i(k)),
\end{equation} where $\boldsymbol{\theta}_i(0)$ is given, $\alpha_i(k)$ denotes the positive step size for agent $i$ satisfying $\sum_{k=0}^\infty \alpha_i(k) = \infty$ and $\sum_{k=0}^\infty \alpha_i^2(k) < \infty$. For a constant matrix $\mathbf{\hat X}_i$, we refer to \cite{hao2025optimalcontrolnonlinearsystems} for a detailed convergence analysis of the update rules in \eqref{lmn}-\eqref{lmn2}.

To sum up, we propose the following algorithm, which is referred to as distributed deep Koopman learning using partial observations (DDKL-PO): 
\begin{algorithm}
\caption{Distributed Deep Koopman Learning using Partial Observations (DDKL-PO)}\label{algorithm:DDKC}
\begin{algorithmic}[1]
 \renewcommand{\algorithmicrequire}{\textbf{Input:}}
 \renewcommand{\algorithmicensure}{\textbf{Initialization:}}
 \REQUIRE Observed partial states trajectory $\boldsymbol{\xi}_i$ in \eqref{eq_batch}.
 \ENSURE  Each agent $i$ builds DNN $\boldsymbol{g}_i(\cdot, \boldsymbol{\theta}_i): \mathbb{R}^{n_i} \rightarrow \mathbb{R}^{n}$. Set iteration index $k=0$, step size sequence $\{\alpha_i(k)\}_{k=0}^K$, and the initial estimated full state matrix $\mathbf{\hat X}_i(0)=C_i^\dagger\mathbf{\bar Y}_i$. \FOR {$k\leftarrow 0$ to $K$}
  \STATE Update $\mathbf{\hat X}_i(k)$ according to \eqref{eq_Xmat_update}.
  \STATE Update $[A_i(k), B_i(k)]$ and $H_i(k)$ following \eqref{lmn} and \eqref{lmn1} using $\mathbf{\hat X}_i(k+1)$, respectively.
  \STATE Update $\boldsymbol{\theta}_i(k)$ following \eqref{lmn2}.
  \ENDFOR
  \STATE Save the resulting $A_i(K),B_i(K),H_i(K),\boldsymbol{\theta}_i(K)$.
 \end{algorithmic} 
 \end{algorithm}

\section{Numerical Simulations}\label{results}
In this section, we apply the proposed algorithm to identify the dynamics of a simple Lunar Lander example from OpenAI Gym \cite{brockman2016openai}, which has a $6$-dimensional system state and a $2$-dimensional control input. The training data are generated using control inputs sampled uniformly from the range $[-1,-1]'$ and $[1,1]'$ with a fixed interval of $0.1$ seconds. 

\begin{figure}[ht]
    \centering
    \includegraphics[width=0.38\textwidth]{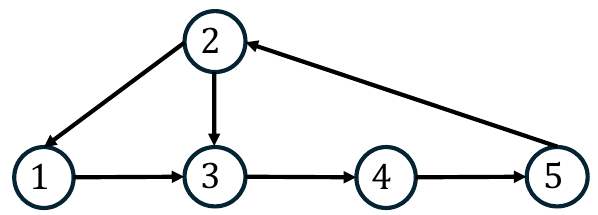}
    \caption{Five-agent directed network with self-arcs, where self-arcs are omitted for simplicity.}
    \label{fig:MAS}
\end{figure}

\textbf{Experiment Setup.} We consider a MAS in Fig.~\ref{fig:MAS},  where the observation matrices $C_i$ regarding each agent $i$ are defined as: \begin{equation}
    \begin{aligned}
        C_1 &= \begin{bmatrix}
   4/7& 3/7&0&0&0&0
\end{bmatrix},\\ C_2 &= \begin{bmatrix}    0&1/2&1/4&0&1/4&0\\
   0 &0&3/5&0&1/5&1/5
\end{bmatrix},\\ C_3 &= \begin{bmatrix}
   0& 1/3&0&1&0&0
\end{bmatrix}, \\ C_4 &=  \begin{bmatrix}
   0& 1/5&1/5&0&3/5&0
\end{bmatrix}, \\ C_5 &=  \begin{bmatrix}
   2/7& 1&0&0&0&3/7
\end{bmatrix}. \nonumber
    \end{aligned}
\end{equation}
For all agents in the MAS, the DNN observable function $\boldsymbol{g}_i(\cdot,\boldsymbol{\theta}_i): \mathbb{R}^{n_i} \rightarrow \mathbb{R}^{12}$ is implemented with three hidden layers of $100$, $64$, and $32$ nodes, respectively, using the $ReLU$ activation function. The DNN is trained using the \emph{Adam} optimizer \cite{kingma2014adam} with a \emph{learning rate} of $10^{-5}$ and a \emph{weight decay rate} of $10^{-8}$. The proposed method is evaluated against the centralized deep Koopman operator (DKO) \cite{dk}, which serves as the benchmark and relies on fully observed states to achieve \eqref{eq_DKO}. For the centralized DKO, a DNN $\boldsymbol{g}(\cdot,\boldsymbol{\theta}): \mathbb{R}^6 \rightarrow \mathbb{R}^{12}$ with the same architecture, training data, and training parameters as the proposed method is employed.

\textbf{Evaluation Metrics.} The estimation error of the proposed DDKL-PO method for each agent $i$ is defined as $$E_{\textrm{DDKL-PO}} = \frac{1}{5}\sum_{i=1}^5\parallel \boldsymbol{\hat x}_{i, t+1} - \boldsymbol{x}_{t+1}\parallel,$$ where $\boldsymbol{x}_{t+1}$ is the constant true state, $\boldsymbol{\hat x}_{i, t+1}$ is obtained from \eqref{eq_DDKO} using $A_i,B_i,H_i,\boldsymbol{\theta}_i$ learned by DDKL-PO in Algorithm~\ref{algorithm:DDKC} based on given $(\boldsymbol{y}_{i, t}, \boldsymbol{u}_t)$. The estimation error of the centralized DKO method is defined as $$E_{\textrm{DKO}} = \parallel \boldsymbol{\hat x}_{t+1} - \boldsymbol{x}_{t+1}\parallel,$$ where $\boldsymbol{\hat x}_{t+1}$ is the predicted state from the centralized DKO in \eqref{eq_DKO} based on given $(\boldsymbol{x}_t, \boldsymbol{u}_t)$. To mitigate the effect of randomness in DNN training, all experiments are repeated five times.

\begin{figure}[ht]
    \centering
    \includegraphics[width=0.42\textwidth]{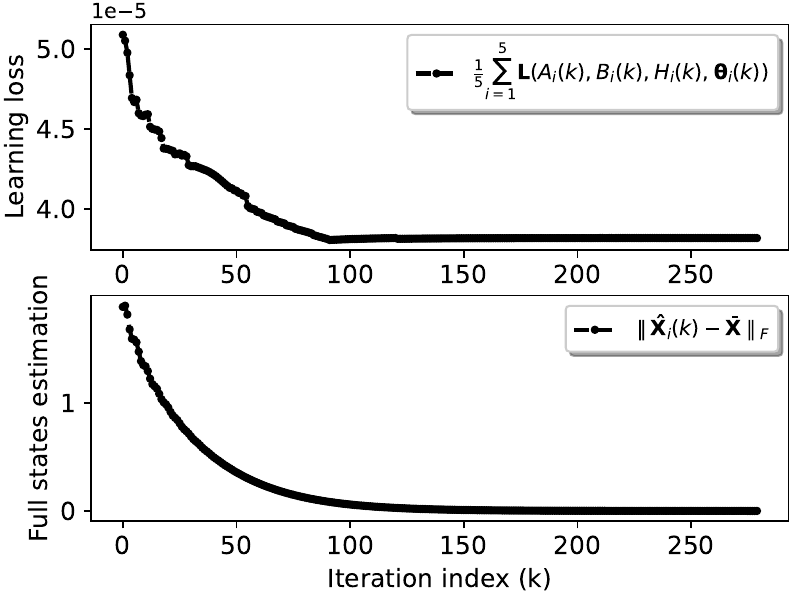}
    \caption{Learning loss in \eqref{eq_loss_compact} and estimation errors in \eqref{eq_Xmat_update}.}
    \label{fig:ddkl_po_learning}
\end{figure}

\begin{figure}[ht]
    \centering
    \includegraphics[width=0.42\textwidth]{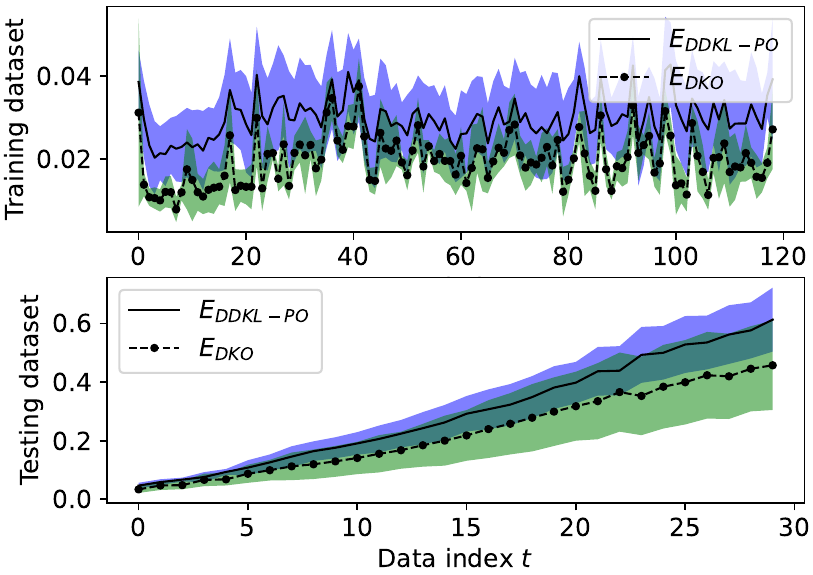}
    \caption{Estimation errors on the training dataset (top) and testing dataset (bottom). The solid line denotes the mean error over five runs, and the shaded area indicates the standard deviation. Results are shown for DKO (green) and DDKL-PO (blue).}
    \label{fig:ddkl_po_testing}
\end{figure}

\textbf{Results Analysis.} Fig. \ref{fig:ddkl_po_learning} illustrates the training process of the proposed DDKL-PO algorithm, showing both the decrease of the loss function in \eqref{eq_loss_compact} and the reduction of estimation errors on the full system state from the update rule in \eqref{eq_Xmat_update}. As shown in Fig. \ref{fig:ddkl_po_testing}, the slightly higher error of DDKL-PO comes from the distributed nature of the partial observations across agents, in contrast to the centralized DKO. Nevertheless, each agent, with only partial measurements, achieves estimation errors comparable to those obtained by the centralized DKO, which leverages full state information.

\section{Concluding Remarks}\label{conclusion}
In this work, we have described a distributed dynamics learning algorithm, termed distributed deep Koopman learning using partial observations (DDKL-PO), for approximating the dynamics of nonlinear time-invariant systems in multi-agent networks with fixed graph structures, where each agent has access only to partial system states. The key contributions of this study include the DDKL-PO problem formulation in \eqref{eq_DDKO} and a simultaneous updating rule for both full-state estimation in \eqref{eq_Xmat_update} and dynamics learning in \eqref{lmn}-\eqref{lmn2}. A key feature of the method is that each agent exchanges only its partial observations and lifted Koopman observables output with neighbors, thereby reducing communication and computational demands while enabling decentralized system identification. Simulation results demonstrated that DDKL-PO attains prediction accuracy comparable to centralized deep Koopman learning, which requires full state information.

A future direction is the development of optimal control strategies based on the learned DDKL-PO dynamics, which remains a promising and challenging research avenue.



\bibliographystyle{unsrt}
\bibliography{hao_refs}

\end{document}